# Links between the Personalities, Styles and Performance in Computer Programming

Zahra Karimi, Ahmad Baraani-Dastjerdi, Nasser Ghasem-Aghaee, Stefan Wagner

**Abstract**

There are repetitive patterns in strategies of manipulating source code. For example, modifying source code before acquiring knowledge of how a code works is a depth-first style whereas reading and understanding before modifying source code is a breadth-first style. To the extent we know, there is no study on the influence of personality on them. The objective of this study is to understand the influence of personality on programming styles. We did a correlational study with 65 programmers at the University of Stuttgart. Academic achievement, programming experience, attitude towards programming, and five personality factors were measured via self-assessed survey. The programming styles were asked for in the survey or mined from the software repositories. Performance in programming was composed of bug-proneness of programmers which was mined from software repositories, the grades they got in a software project course and their estimate of their own programming ability. In a statistical analysis we found that Openness to Experience has a positive association with breadth-first style and Conscientiousness has a positive association with depth-first style. We also found that, in addition to having more programming experience and better academic achievement, the styles of working depth-first and saving coarse-grained revisions improve performance in programming.

**Keywords**: Programming Styles, Personality, Five-Factor Model, Programming Performance, Statistical Analysis

## 1. Introduction

Software development is a human task and individual differences are known to be present in it and have been identified in previous research. For example, developers differ in their performance (Sackman et al. 1968), in the way they make their judgments and take their decisions (Feldt et al. 2010). Programmers also differ in their working habits, in the way they generate and comprehend source code (Cox and Fisher 2009). In other words, each programmer might use a different strategy in modifying source code.

Cox and Fisher (2009) used the term *programming styles* to describe recurring strategies in manipulating source code. For example, programming can be done *top-down* and or *bottom-up*. In *top-down* programming, more tasks might be done in model space and in *bottom-up* programming, more tasks might be done in code space.

*Programming styles* have been studied for several decades. The researchers believed that these styles are the result of programming experience. Therefore, finding them would help train novices. For example, Vessey (1985) studied think-aloud data of 4 novices and 4 experts while they were debugging a piece of Cobol code. He found that experts' style is systematic (*breadth-first*) but novices' style is opportunistic (*depth-first*). Ko and Uttl (2003) also observed 75 students at debugging. They confirmed that students who work *breadth-first* were more experienced. Mayrhauser and Vans (1997) studied think-aloud data of 4 professionals and found that programmers frequently switched between solution space (*bottom-up*) and problem space (*top-down*); however, language expertise encouraged *bottom-up* and domain knowledge encouraged *top-down* styles.

Cox and Fisher (2009) reviewed *programming styles* and hypothesized that, in addition to experience, other individual factors may explain these differences. They themselves studied the influence of gender on *top-down* and *bottom-up* styles (Fisher et al. 2006) and reported that males tend to use *top-down* styles while females tend to use *bottom-up* styles. Cox and Fisher (2009) also claimed that personality influences *programming styles*.

Personality refers to individual differences among people in behaviour, cognition and emotion patterns (Michel et al. 2004). There are studies on the influence of personality on individual performance or team work in programming (Cruz et al. 2011 and 2015) but to the extent we know, there are no empirical studies on the influence of personality on *programming styles*. We should mention that Cox and Fisher (2009) themselves did a pilot study on the influence of personality on *programming styles* but did not report the results.

In the present study, we make a first attempt to fill this gap and investigate the influence of personality on *programming styles*. One first objective is to find out which personalities use which *programming styles*. Moreover, we are interested in consequences of relations between personality and *programming styles*. Therefore, we also investigate relations between personality and *programming styles* on one hand and performance on the other.

There are various models for describing personality in psychology. In this study we focus on the Five-Factor Model. This is a comprehensive and well-established model in psychology (Digmann 1990) and has recently been used frequently in software psychology research (Cruz et al. 2011 and 2015). Moreover, Cox and Fisher (2009) recommended the Five-Factor Model in the area of *programming styles*.

We conducted an empirical study and surveyed personality factors of 65 volunteer student programmers in software engineering at the University of Stuttgart. We used indicators in the literature and devised our own questionnaire to survey *programming styles* discussed in Cox and Fisher (2009) study. Moreover, we used a mining algorithm to extract one of those programming styles from repositories of software projects under study. In addition to personality factors and programming styles, we also collected data on age, gender, academic achievement (overall grade of all university exams), prior programming experience and programming performance. On the collected data, we did correlational analyses to see which personality and/or other factors affect programming styles and which programming styles and/or other human factors affect programming performance.

In this paper, we report our results on the links between investigated human factors, programming styles and performance. After explaining the background and related work in section 2, we describe the design of the empirical study in section 3. Then we present the statistical findings in section 4 and discuss them in section 5. Limitations and future work are presented in sections 6 and 7, and we conclude the paper in section 8.

## 2. Background and Related Work
### 2.1. Personality Psychology

Psychology is a collection of scientific methods and theories for understanding human nature, and personality is a part of psychology that considers characterizing individuals. Psychologists usually describe personality by several personality traits. For each individual, each personality trait has a numeric score indicating how much of the trait an individual possesses.

Yet they are not sure how many and which narrow traits should be considered in a comprehensive personality model; generally, psychologists have a good consensus about the comprehensiveness of the Five-Factor Model (Digmann 1990). In this model, five broad and distinct traits describe personality: *Openness to Experience*, *Conscientiousness*, *Extraversion*, *Agreeableness* and *Emotional Stability*.

*Openness to Experience* shows to which extent we are creative, interested in art, willing to try new things and aware of our feelings. *Conscientiousness* shows how much we prefer planned behaviour, self-discipline, acting dutifully and aiming for achievement. *Extraversion* shows how much we engage with the external world, interact with people and assert ourselves. *Agreeableness* shows how humble we are, how much we value getting along with others, how considerate, friendly, generous, helpful we are and to which extent we are willing to share our feelings with others. *Neuroticism* shows how much we experience negative emotions such as anger, anxiety and depression.

Personality tests, which usually are an inventory of multiple-choice questions, operationalize personality traits. This means that these questions identify observable evidence of traits in an assessable way. As an example, Costa and McCrae (1992) developed the NEO-PI-R inventory which allows a comprehensive assessment of the five personality factors to be done. Since the NEO-PI-R inventory is commercial and, therefore, researchers cannot freely use it, Gow et al. (2005) provided the assessment items for the Five-Factor Model in the public domain of personality scales and items (IPIP, Goldberg, 1999). The items in IPIP are validated by psychologists and translated into many languages. Therefore, this pool provides us with what we need: a free, valid, reliable and comprehensive personality test.

In summary, although personality descriptions and tests are developed in the context of psychology research, their research results provide software engineering research with required background and tools, theories and assessments. In this study, we focus on the Five-Factor Model for describing personalities of programmers and the IPIP for assessing personalities.

### 2.2. Programming Styles

Cox and Fisher (2009) explained 4 basic *programming styles*: *top-down*, *bottom-up*, *depth-first* and *breadth-first*. In the *top-down* style, programmers investigate the model space to solve the problem, whereas in the *bottom-up* style, programmers investigate the code space to solve the problem. In the *depth-first* style, programmers proceed with the first alternative to solve the problem and, therefore, they are opportunistic. In the *breadth-first* style, programmers examine all alternatives before proceeding, and they are systematic. Although these styles are a personal preference, it does not mean some people use *top-down* and do not use *bottom-up* or vice-versa (Mayrhauser and Vans 1997) or some people use *depth-first* and do not use *breadth-first* or vice-versa (Cross 2005). In the following section we illustrate these basic styles by an example.

Programmers also differ in the amount of work they do in each session. Cox and Fisher (2009) believe that those who write a large amount of code before bug removal are coarse-grained programmers and others are fine-grained programmers. But since in the modern IDEs all programmers see their syntax bugs immediately, we believe deferring bug removal sessions and in other words saving large amount of work before the next bug-removal session means that these programmers avoid bug removal and keep working on buggy code.

We measured the amount of work in each revision, instead, to assess coarse-grained programmers. Therefore, in addition to 4 basic programming styles, we included 2 other programming styles: *bug-removal-avoidance* and *large-revisions*.

### 2.2.1. Example

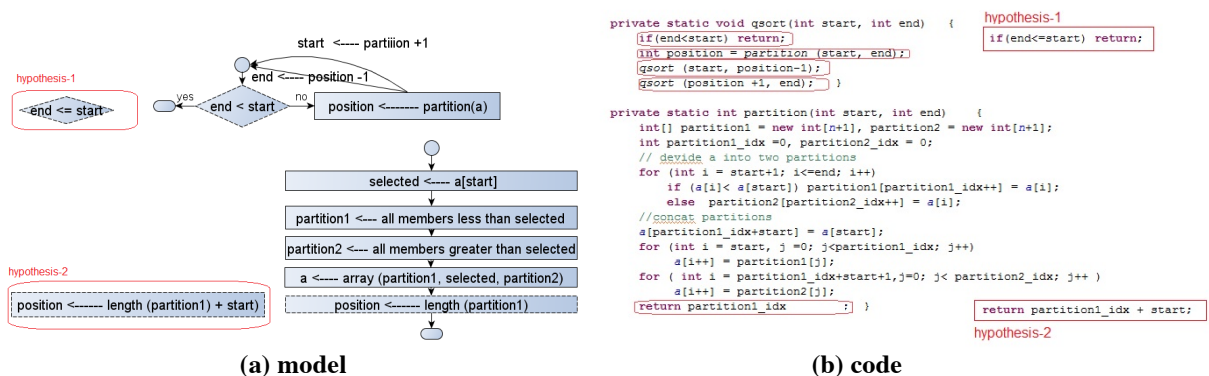

**(a) model**            **(b) code**
**Fig. 1. A sample model (and equivalent code) of quick sort with an infinite loop bug**

Suppose a programmer is assigned to modify a quick-sort code which has an infinite loop bug. Figure 1 shows the model and code of this buggy quick sort. The programmer might make two hypotheses. The first hypothesis is: "a sequence of length 1 does not need sorting". The second and correct hypothesis is "the partition method returns a wrong value." The programmer might use different flows in making hypotheses and evaluating them. Figure 2 shows possible flow diagrams.

The two top diagrams (a and b) in figure 2 show *top-down* work and the two bottom diagrams (c and d) show *bottom-up* work. The two diagrams on the left (a and c) show *depth-first* and the two diagrams on the right show (b and d) *breadth-first* work. As the horizontal swimlanes in figure 2 show, there are three layers: "model", "code" and "run" and there are four basic actions in this debugging: "find hypothesis", "modify model", "modify code" and "test code".

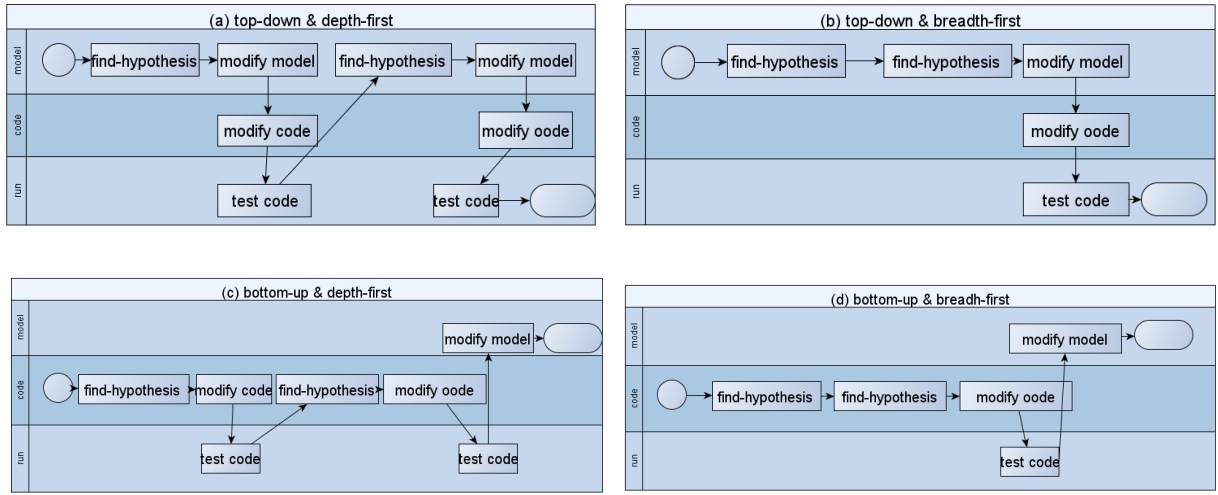

**Fig. 2. Sample flows of programming styles in debugging**

In working *top-down* (figure 2 a and b), a programmer traces conditions, calls and return parts of the model to find a hypothesis. He first modifies the model and then the code. However, in working *bottom-up* (figure 2 c and d ), a programmer traces code and finds conditions, calls and return parts to make a hypothesis. He first modifies the code and then the model. As figure 2 shows, in *top-down* work more actions are done in the model layer, whereas in *bottom-up* work more actions are done in the code layer.

Working *depth-first* (figure 2 a and c), a programmer tests the first hypothesis he finds and, if it does not solve the problem, creates another hypothesis. However, working *breadth-first* (figure 2 b and d), a programmer does not commit to the first hypothesis and keeps investigating hypotheses before modifying and testing. As figure 2 shows, in *depth-first* work more actions are done in the depth whereas in *breadth-first* work more actions are done in the breadth.

**2.2.2. Theory**

Cox and Fisher (2009) presented a conceptual framework (see figure 3) and explained that, as a consequence of different situations, tasks and individual factors, programmers prefer to use certain *programming styles*. Situation is a set of factors that represent the characteristics of the surrounding, where, how, when and with whom the task is performed. Time available, organizational culture and teams are examples of situational factors. Technical and programming-relevant characteristics are task factors such as difficulty and functionality. Individual factors represent all personal factors, whether they are unchangeable such as personality (see internal component in figure 3) or changeable such as experience (see external component in figure 3). In this paper, we focused on individual factors to study programming styles.

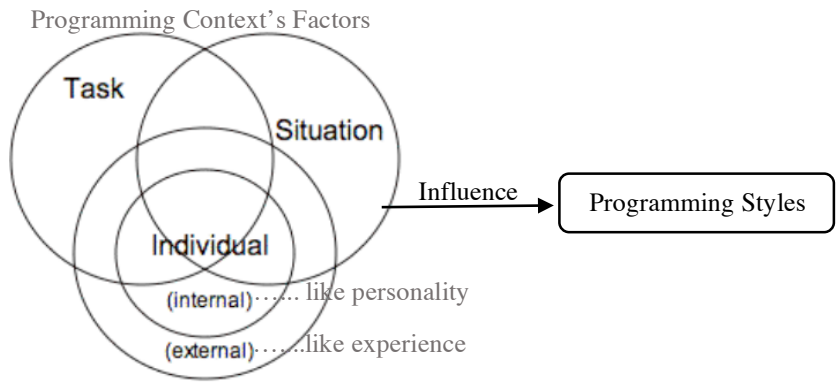

**Fig 3. Programming Context influences on Programming Styles, derived from Cox and Fisher (2009)**

## 2.2.3. Literature Review

**Table 1. Summary of the literature on individual differences and programming styles**

| paper | individual factors | performance | results |
|---|---|---|---|
| Vessey (1985) | experience | debugging score | experts use *breadth-first* <br> novices use *depth-first* <br> experts use debugging score |
| Webb et al. (1986) | biography <br> aptitude <br> cognitive styles | test score | *top-down* increases test score in generating new code <br> *bottom-up* increases test score in modifying existing code <br> Students mainly use *bottom-up* |
| Mayrhauser and Vans (1997) | domain knowledge <br> language expertise | | domain knowledge encourages *top-down* <br> language expertise encourages *bottom-up* |
| Ko and Uttl (2003) | age, gender, major <br> domain knowledge <br> language expertise <br> attitude | debugging score | Computer science major encourages *breadth-first* <br> experience encourages *breadth-first* <br> attitude encourages *breadth-first* <br> *breadth-first* increases debugging score |
| Fisher et al. (2006) | gender <br> Spatial cognitive ability | test score | males use *top-down* <br> females use *bottom-up* <br> *bottom-up* increases test score |

In this section, we review the empirical literature on the individual differences in programming styles (see Table 1). Vessey (1985) conducted a protocol analysis on the think-aloud data of 4 novice and 4 expert programmers in debugging the same Cobol code. He extracted problem solving strategies and found that expert programmers tended to use *breadth-first*; they kept reading code and comments to understand it. They did not commit to their first hypothesis and were not blind to new information. However, novices were more probable to use *depth-first*; they executed the code sooner and persisted with their first hypothesis.

Webb et al. (1986) studied 30 novice students (11–14 years old) who were writing a simple BASIC program. They used taped records and prints of computer verbatim and then coded the programming behaviour. They measured the frequency of coded variables to recognize their tendency to use *top-down* or *bottom-up*. In their study, an abstract plan such as a design plan before writing code indicates *top-down* and a concrete plan such as an operation plan indicates *bottom-up*. For individual differences, they surveyed biography variables, aptitude and the cognitive style of students. They performed a correlational analysis on all variables and could not find consistent significant relations between individual differences and programming styles and did not report them. But they reported that students mainly used *bottom-up* style. Moreover, they found that the degree of *top-down* style affected the learning outcome of students in generating new code and the degree of their *bottom-up* style affected their learning outcome in modifying existing code.

Mayrhauser and Vans (1997) did a field observation and recorded think-aloud data of 4 professionals during corrective maintenance of a large-scale software. In protocol analysis, they coded and enumerated hypotheses of programmers to see whether programmers tend to produce domain-level hypotheses (*top-down*), program-level hypotheses (*bottom-up*) or situation-level hypotheses (switch between *top-down* and *bottom-up*). They found that programmers frequently switched between *top-down* and *bottom-up*; however, programmers with domain knowledge tended to use *top-down* and programmers with language expertise tended to use *bottom-up* styles.

Ko and Uttl (2003) observed 75 undergraduate computer science, psychology and statistics students who were debugging a simple program in an unfamiliar system (programmable statistical package). In addition to think-aloud data, they analyzed and coded video and screen recordings to find comprehension behaviours. They used cluster analysis and classified behaviours into 3 distinct clusters: *depth-first*, *inactive* and *breadth-first*. They investigated the mean differences of age, gender, domain knowledge, programming experience, attitude and major in these clusters. They found that students who tended to use *breadth-first* were generally more experienced, had a positive attitude towards experiments and were computer science students. Moreover, students who tended to be *inactive* were psychology students and had less programming experience.

Fisher et al. (2006) studied 30 graduate and undergraduate students performing a set of maintenance tasks on Java. Participants did some post-maintenance tasks on how well they remembered the size, location and name of methods. Moreover, the participants filled out spatial cognition tests, location memory, object memory and mental rotations tests. Fisher et al. (2006) correlated all the measures, the scores of spatial cognition tests, post-maintenance tasks and maintenance tasks for males and females separately. They found that males tended to use mental rotation ability (highly abstract) and remembered the size and functionality of methods. But females tended to use location memory (highly concrete) and remembered the name and location of methods. Fisher et al. (2006) concluded that males may use *top-down* and females *bottom-up* styles. Moreover, they found that programmers who used *bottom-up* achieved better test scores.

As we summarized in Table 1, there are empirical studies on the influence of experience, gender, attitude and mental ability on *programming styles* but as far as we know, there is no empirical study on the influence of personality on *programming styles*. In this study, we fill this gap. To do so, we rely on Cox and Fisher's theory and investigate the influence of personality factors on *programming styles*.

## 3. Study Design

### 3.1. Research Questions

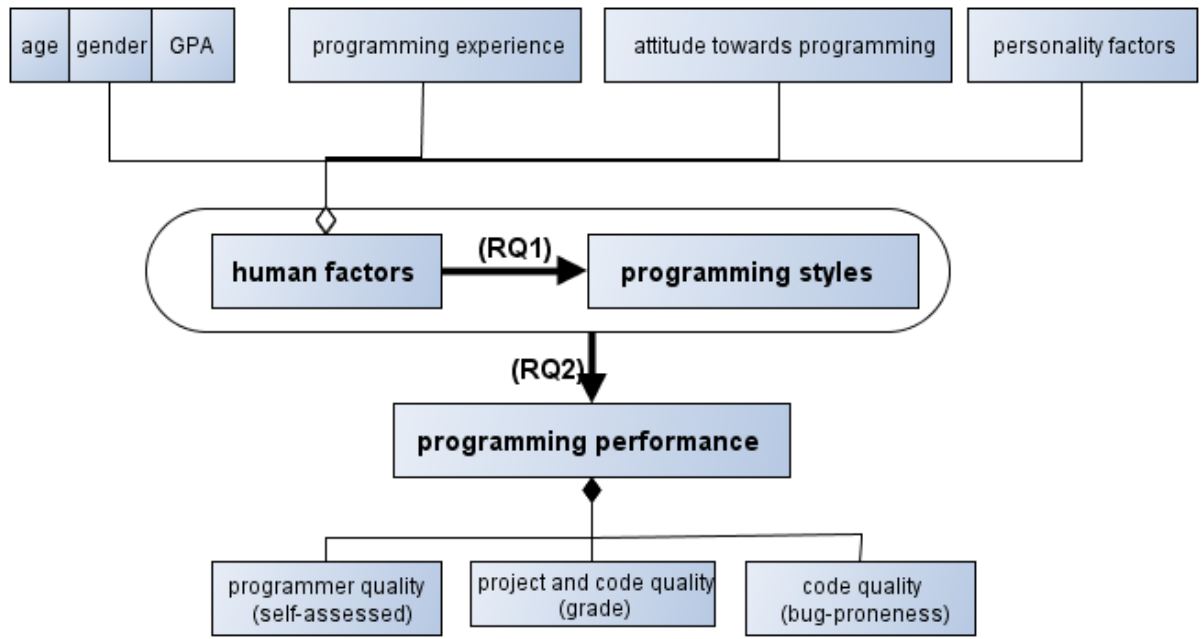

**Fig. 4 The structure of research questions**

We showed the structure of research questions in figure 4. First, we investigate the influential factors on *programming styles* to see how strongly personality affects *programming styles*. Then we study the effect of all factors on performance to empirically figure out the potential benefits of *programming styles*. In summary, we answer two research questions:

RQ1. Which human factors affect certain programming styles?

RQ2. Which factors, human factors and programming styles affect performance?

As figure 4 shows, in addition to personality factors we study age, gender, academic achievement, programming experience, and attitude towards programming. To measure programming performance, we use indicators of programmer quality, project quality and code quality.

### 3.2. Target population

The target group for the study are German software engineering students mostly at the University of Stuttgart in the first and advanced software project courses (in 2012 and 2013). In their first software project course ("Sopra"), second-year students work for one semester in 3-member teams to develop the same real software project. In this course, three expert software engineers (one in the role of the customer and two as supervisors) evaluate the students' work (system design and system implementation) and each gives a score to the whole team. The average score determines the state of fail or pass in the course. In advanced software project courses ("Studienprojekt" or "Stupro") students work for 12 months in teams of 6 to 12 programmers and develop a real software project. In addition to the quality of the whole project, supervisors examine the contributed code of each student (and other factors such as their motivation) and give individual scores to each team member.

The initial idea for finding volunteers was presenting the study and giving information about our research. It became apparent that interested students needed more privacy and time to think and react. Therefore, we developed a brochure for introducing the study and attracting the volunteers.

We first talked with the course tutors (and the instructor of the course) and gave the brochure to them. After getting the initial agreement, we distributed the brochure among students directly or indirectly. Volunteers could signal their interest on the website of the course or send e-mails. They were supposed to fill in an on-line questionnaire and give us read-only access to the repositories of their source code. Moreover, they were supposed to write appropriate descriptions for their commits in the repository.

In December 2012, in the middle of the SOPRA-2012 course, we had three volunteering teams two of which ultimately did not participate in the study. One of the teams did not finish the course and one of them cancelled their participation. In February 2013 and at the end of the course, we attracted another volunteering team.

In June 2013 we had one 7-member volunteering team in an MSc. course out of which 6 students participated in the study. We also attracted 4 volunteering teams who had just finished advanced software project courses out of which 19 developers participated in the study. Moreover, we had one volunteering team external to the University of Stuttgart out of which 3 people participated in the study.

In November 2013, in the middle of the SOPRA-2013 course, we had 13 three-member volunteering teams two of which ultimately did not participate in the study. One team gave up the course and another one failed to pass it. One of the teams also participated with only one member.

In summary, 8 different software projects, 19 teams and 65 volunteers participated in the study. We had 18 repositories: 8 git (http://git-scm.com/), 1 mercurial (http://mercurial.selenic.com/) and 9 SVN (Collins-Sussman et al. 2011). We excluded one of the software repositories because it only had one commit. All volunteers were developing real software projects in Java (57) and/or C# (8). 13 teams and 37 volunteers with 2 different software projects were in the second year of their studies. 4 teams and 19 volunteers were in the third or fourth year of their studies. 1 team and 6 volunteers were Master of Science students. 1 team and 3 volunteers were graduates who were maintaining their software project.

### 3.3. Data collection procedure
We used a self-assessed survey and repository mining to collect data except for the grades students achieved in the software project courses. We collected some of the grades from the volunteers themselves (by e-mails, 5 grades were missing) but we got the other grades from the supervisors since some students were not aware of their grades and only get fail or pass votes. In the following, we will explain the design, pre-testing and administration of the questionnaire items as well as the repository mining procedure.

### 3.3.1 Design and pre-testing of questionnaire

The questionnaire consisted of 4 main parts: (1) personality items, (2) programming style indicators, (3) experience, ability and background items and (4) attitude items (see items in the appendix).

(1) Personality items: Since the data collection site was Germany, we used the German translation of IPIP-50 (validated by Gow et al. 2005 and translated by Streib in the Research Center for Biographical Studies in Contemporary Religion at the Universität Bielefeld) to measure the Five Factors. In this questionnaire each personality factor is assessed by 10 short descriptions by which the respondent should decide how strongly he/she agrees with the item. To do so, the respondent selects one value in the range of 1 to 5 (1–5 Likert scale). Some of the items indicate high values in personality factors and some indicate low values. For example, strong agreement with "Spend time reflecting on things" indicates high *Openness to Experience* and agreement with "Am not interested in abstract ideas" indicates low *Openness to Experience*. The items in the questionnaire, number of positive or negative items, wording and even the order of the items were psychologically validated. Therefore, the questionnaire is psychologically standardized in a way that decreases the level of misunderstanding and misleading items to a minimum. However, we used the mother-tongue for the personality questionnaire and also checked the selected questionnaire and administration procedure with the psychology advisor of the research, Jan-Paul Leuteritz. We also rechecked the on-line questionnaire with the help of a language expert to be sure it did not have any copy-paste problems.

(2) Programming style indicators: We used Cox and Fisher (2009) as the theoretical framework of our work and, therefore, included the programming styles they introduced in the empirically study. In the following we explain them:

- *Top-down and bottom-up* styles: We used Fisher et al. (2006) indicators in code-comprehension to assess these styles: *top-down* programmers usually investigate the length and functionality of a method and *bottom-up* programmers look into names and locations inside the code.

- *Breadth-first and depth-first* styles: We used Vessey (1985) indicators in debugging to assess these styles: *breadth-first* programmers examine all possible ways before modifying the code, while *depth-first* programmers apply the first hypothesis they found.

- *Bug-removal-avoidance style*: We used the Cox and Fisher (2009) indicator in syntax-bug-removal to measure *bug-removal-avoidance* style. A large amount of work before the next bug-removal session indicates *bug-removal-avoidance*.

(3) Experience, ability and background items: To measure experience, we asked the participants about the years/months of prior programming experience including school and university work, the programming languages they worked with and their largest code contribution. Moreover, we asked about years/months of professional programming. Experience items were an adaptation of the previously used survey for newly enrolled students of the software quality course at the University of Stuttgart.

To measure ability, we asked for an estimate of their own ability in programming and their own ability in comparison with their friends in the team (like Newsted 1975). We also added some background questions about age, gender, GPA and the language of participants.

(4) Attitude items: We adapted 10 items of the Computer Science Attitude Survey (Wiebe et al. 2003) which was derived from Fennema-Sherman mathematics attitudes scales (Fennema and Sherman 1976). We had 3 items for confidence-in-learning-programming, 2 items for attitude-towards-success-in-programming, 2 items for computer-science-as-a-male-domain, 1 item for usefulness-of-programming and 2 items for effective-motivation-in-programming. Like personality items, attitude items were also in 1-5 Likert scale. We added attitude items in the middle of the study and we don't have attitude values for all participants.

To make sure that our questionnaire was clear and appropriate for the participants, we piloted it two times among six expert software engineers who knew the student respondents and worked at the Institute of Software Technology at the University of Stuttgart. They were at least PhD students in the field of empirical software engineering and had a good sense of the research objectives and approach. We also had the questionnaire checked by a language expert.

The personality part of the questionnaire was in German, while other parts were in English. This was because the first author does not speak German. We published our survey in the e-learning platform of the University of Stuttgart: https://ilias3.uni-stuttgart.de/goto.php?target=svy_497145&client_id=Uni_Stuttgart

### 3.3.2. Administration of the questionnaires

We communicated with the volunteers by e-mails and arranged a meeting for filling out the survey. We met them in groups of 1 to 5 students and gave the survey link to them. Some of the participants filled the survey out on their own laptops. Several participants were not available at the time of data collection and filled it out remotely.

Before survey administration, we emphasized that their answers would be kept confidential and every personal style has its own beauty. We wanted them to describe themselves as they are now, not as they wish to be in the future, and asked them to see themselves honestly, in relation to other people they know of the same sex, and roughly their age. Finally, we informed them that there was no time limit. They took 20 minutes on average.

Respondents could get support for the meaning of items if it was not clear to them but almost all the respondents filled out the questionnaire without help. Several students asked how to judge themselves; for example, one of them asked whether scripting counted as programming experience. Only one of the participants had problems with understanding the German words of the personality items. We provided him with a hard copy of the English equivalent.

After filling out the survey, the participants signed an agreement which allowed us to save, analyze and publish their personal data anonymously and received a small compensation (10 Euros). We distributed the survey in June/July 2013 (34 respondents) and December 2013 (37 respondents).

### 3.3.3. Repository mining

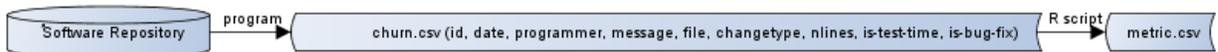

**Fig. 5 Data extraction procedure from software repositories**

We wrote a Java program to extract *churns* from software repositories (see Fig. 5). Each *churn* shows the programming work-unit on one revision and one file. As figure 5 shows, *churn*s is a dataset with these columns: *revision-id*, *commit-date*, *programmer*, *message*, *source-file*, *change-type*, *nlines, is-test-time* and *is-bugfix*. *Programmer* column indicates the login name of the programmer who committed the code. *Change-type* is the type of change on the file: created, deleted, modified, and so on. *nlines* is the total number of lines of code changed in the unified format[1]. The unified format is the format software repositories use for comparing file revisions. In this format, lines added are preceded by "+", lines deleted are preceded by "-"and modifications are modeled by several lines added and deleted. The column *is-test-time* indicates whether the revision was saved during test time or development time. The last column, *is-bugfix*, indicates whether a bug fix in the revision or something else has been done.

We wrote R scripts (Statistical Computing with R) to extract these metrics from the *churns* dataset: *file-count-in-revision*, *line-count-in-revision* and *bug-ratio*. Although volunteers were supposed to use the repository, one 3-member team failed to use it appropriately; also, we could not map one of the participants to the repository logs. Therefore, we missed 4 participants in repository styles (n = 61).

input: programmer code (p)
output: mean file count in revisions of programmer p
file-count-in-revision = select mean
    (select Count(source-file)
    from churn
    where programmer = p
    and change-type = "modified"
    group by revision.id)

**(a) file-count-in-revision**

input: programmer code (p)
output: mean line count in revisions of programmer p
line-count-in-revision = select mean
    (select sum(nLoC)
    from churn
    where programmer = p
    and change-type = "modified"
    group by revision.id)

**(b) line-count-in-revision**

input: programmer code, p
output: modified files by p in develop time
modified.files = select unique(source-file)
    from churn
    where programmer = p
    and change-type = "modified"
    and is-test-time = NO

**(c) modified-files**

input: -
output: modified files in bug-fix revisions of test time
buggy.files = select unique(source-file)
    from churn
    where is-test-time = YES
    and is-bug-fix = YES

**(d) buggy-files**

[1] http://en.wikipedia.org/wiki/Diff_utility#Unified_format

**Fig. 5 Pseudo scripts for getting repository metrics**

*large-revisions*: We computed the mean Lines of Code and mean Number of Files in revisions for each programmer to assess tendency to *large-revisions*. For simplicity, we focused only on modifications and discarded other change types. Figure 5a and 5b shows the pseudo scripts. As figures 5a and 5b show, we first counted lines (files) in each revision and then computed mean of counts.

*bug-ratio*: We measured *bug-ratio* to estimate the code quality of each developer. Bell et al. (2013) introduced this metric to study bug-proneness of programmers in software repositories. The *bug-ratio* indicates in which percentage of source files a programmer worked on during development, bugs are found during test. A large *bug-ratio* shows that a programmer is bug-prone. Bell et al. (2013) show that *bug-ratio* is relatively constant in different releases and, therefore, can be used as a performance criterion.

We considered the first 80% of the work as the development phase and the remaining 20% as the testing phase. Moreover, we considered all files changed during testing with bug-fix commits as buggy files (like Bell et al. 2013). To fill in the *is-bugfix* column we used a Java program and searched all key words, bug, bugfix, fix, bug fix and etc. We then manually checked the churn dataset and modified it. Although students were supposed to write a meaningful description for each commit, one 3-member team failed to add any descriptions and we missed 3 more participants in this metric (n = 58). As figures 5c and 5d show, we first computed "modified-files" and "buggy-files" and then computed *bug-ratio* using formula 1.

**Formula 1- bug.ratio formula**

$$bug.ratio = \frac{count \cap (modified.files , \ buggy.files)}{count(modified.files \ )}$$

## 3.4. Variables

**Table 2. Summary of variables in the study**

| Personality factors | Openness to Experience (5-50), Conscientiousness (5-50), Extraversion (5-50), Agreeableness (5-50), Emotional Stability (5-50) |
|---|---|
| programming experience | years of programming experience (5-point scale), years of professional experience (4-point scale), number of programming languages , largest code contribution (5-point scale) |
| attitude towards programming | confidence-in-learning-programming (3-15), attitude-towards-success-in-programming (2-10), effective-motivation-in-programming (2-10), attitude-towards-usefulness-of-programming (1-5), computer-science-as-a-male-domain(2-10) |
| programming styles | top-down.1 (0, 1), top-down.2 (0, 1), bottom-up.1 (0, 1), bottom-up.2 (0, 1), breadth-first.1 (1-5), breadth-first.2 (1-5), depth-first.1 (1-5), depth-first.2 (1-5), bug-removal-avoidance (0, 1), large-revisions-file-count, large-revisions-line-count |
| background | age, gender (1-2), GPA (3-point scale), years-at-university (1-3) |
| performance | performance (factor score) |

We used the collected data to prepare the following variables for data analysis (see Table 2 for summary of variables).

- Personality factors: As we already mentioned, there are 10 responses in the Likert 1-5 scale for each personality factor. Some responses influence the score positively and some negatively. Following computation rules in the personality questionnaire, we reverted the negative ones (6- response value) and summed the responses to compute a personality score, 5-50, for each factor. Each participant has 5 personality scores and each score is between 5 and 50.

- Experience: We used the questionnaire items directly as experience indicators. For example, regarding professional programming experience, there is a 4-point scale: less than 6 months, between 6 months and 1 year, between 1 and 3 years and more than 3 years.
- GPA: We collected GPA in a 3-point scale in the questionnaire.
- Attitude towards programming: As we explained before, we collected 5 attitude scales. We used the same procedure in personality factors to compute scores of attitude scales.
- Programming styles: In some styles, there is more than one indicator. For example, in *bottom-up* style, there are two indicators, name-orientation and location-orientation. In summary, each person has 11 scores for 6 programming styles.
- Performance: There are 3 performance variables: (1) Grade: In academic grading in Germany, they use a 1-5 grading scale to evaluate performance of students. (2) Programmer ability: We collected it from the questionnaire (3) bug-ratio: We computed it in repository mining. We did Principle Component factor analysis to analyse (??) factors among our variables. Factor analysis is a statistical method which identifies a potentially lower number of unobserved variables called factors from observed, correlated variables. We found that all three variables reflect only one factor. To find each participant's placement on this factor, we used the Regression Scores technique and computed factor scores (DiStefano et al. 2009).

### 3.5. Analysis Procedure

To answer the research questions, we used correlation analysis using Pearson's correlation coefficient and entered all significant variables into linear regression. We repeated the regression analysis for all subsets of variables and selected the best one with lowest error and significant coefficients. We used R (Statistical Computing with R) to do the regression and correlational analyses. The significance level in all analysis is 0.05.

We should mention that since we added attitude items in the middle of the study, we missed attitude scores for 34 participants. Therefore, we did not enter attitude scores into regression analysis. We only report and discuss significant correlational results in a separate section.

## 4. Results

Of the 65 participants, 7 were female and 58 were male. 16 participants were 19 – 20 years old, 28 participants were 21 – 23 years old and 21 participants were 24 years old and above. Nearly

57% (37) of the participants were in the second year of their studies. Almost all participants (except 2) used German as their first language in which they do most of their thinking, creating and writing.

Almost all participants (except 2) had at least 1 year of programming experience including school and university work and 67 % of participants (44) had more than 3 years of programming experience. Moreover, 15 participants had at least 6 months of professional programming experience. None of the participants had more than 3 years of professional programming experience. None of the participants was a beginner in Java and nearly 77 % (50) of the participants had made code contributions in at least 3 programming languages. The largest code contribution for more than 80 % of participants (55) was at least 1 KLOC. The largest code contribution for 34 participants was at least 5 KLOC.

More than half of all participants (35) liked to earn money with programming as a full-time work and other participants (except 5) liked it but prefer other software engineering tasks like analysing, designing or testing. Therefore, this sample represented intermediate German student programmers and prospective professional software developers.

## 4.1. RQ1. Which human factors affect programming styles?

**Table 3a –Significant correlations between human factors and programming styles**

| factor | style | n | r |
|---|---|---|---|
| Conscientiousness | depth.first.1 | 65 | 0.252* |
| attitude towards usefulness of programming | depth.first.2 | 31 | 0.426* |
| Openness to Experience | breadth.first.1 | 65 | 0.247* |
| Openness to Experience | breadth.first.2 | 65 | 0.322* |
| attitude towards success in programming | breadth.first.1 | 31 | 0.406* |
| years at university | bottom.up.1 | 65 | -0.271* |
| years of programming experience | bottom.up.2 | 65 | 0.297* |
| attitude towards usefulness of programming | bottom.up.2 | 31 | 0.368* |
| age | bug avoidance | 65 | 0.255* |
| gender | bug avoidance | 65 | -0.284 |
| age | number of files | 61 | 0.298* |
| years at university | lines of code | 61 | 0.310* |
| years at university | number of files | 61 | 0.415* |
| years of professional experience | lines of code | 61 | 0.257* |

\*: $p < 0.05$, \*\*: $p < 0.01$

**Table 3b –Regression equations between programming styles and human factors**

| style | regression equation | R square | degree of freedom |
|---|---|---|---|
| depth first 1 | 0.049* Conscientiousness +1.446 | 0.063 | 63 |
| breadth first 1 | 0.035* Openness to Experience +2.655 | 0.061 | 63 |
| breadth first 2 | 0.060* Openness to Experience +0.645 | 0.104 | 63 |
| bottom up 1 | -0.175*years at university+0.598 | 0.073 | 63 |
| bottom up2 | 0.094*years of programming experience+0.438 | 0.088 | 63 |
| bug removal avoidance | -0.448*gender+0.896 | 0.080 | 63 |
| large revisions. line count in revision | 76.84*years at university+18.32 | 0.096 | 59 |
| large revisions. file count in revision | 0.839*years at university+1.453 | 0.171 | 59 |

Table 3 shows the answer to RQ1. Table 3a shows significant correlations between human factors and *programming styles* we studied. Table 3b shows the results of regression analysis. In summary, table 3 indicates that:

- Males avoided bug-removal more than females did.
- Programmers with more programming experience used *bottom-up* style.
- Programmers with higher years at university tended to save larger revisions and did not tend to use *bottom-up* style
- *Conscientiousness* and *Openness to Experience* were the evident personality factors in the *programming styles* we studied.

  - *High Openness to Experience* programmers used *breadth-first* style more often than *low Openness to Experience* programmers.

  - *High Conscientiousness* programmers used *depth-first* style more often than *low Conscientiousness* programmers.

- Programmers with a positive attitude towards programming used *depth-first*, *breadth-first* and *bottom-up* styles.

## 4.2. RQ2. Which factors, personality factors and programming styles affect performance?

**Table 4 Significant correlations and regression equation between programming performance and other variables**

| | variable | n | r |
|---|---|---|---|
| programming performance | age | 54 | 0.327* |
| | GPA | 54 | 0.364** |
| | years at university | 54 | 0.535** |
| | years of programming experience | 54 | 0.439** |
| | number of programming languages | 54 | 0.545** |
| | largest code contribution | 54 | 0.386** |
| | attitude towards computer science as a male domain | 25 | 0.400* |
| | depth first 1 | 54 | 0.361* |
| | large revisions line count | 54 | 0.276* |
| | programming performance = 0.249*number.of.programming.languages + 0.307*depth.first.1 + 0.490*GPA + 0.001* largerevisions.line.count + 0.130*age - 3.895 | df =48 | Rseq = 0.588 |

Table 4 shows that:

- Programmers with more programming experience have better performance than those with less programming experience.
- Programmers with higher academic achievement perform better than programmers with lower academic achievement.
- Older programmers are better programmers than younger ones.
- The following *programming styles* affect performance.
  - Programmers who tended to use *depth-first* more often were better programmers than ones who did not tend to use *depth first* less.

o Programmers who saved *larger revisions* were better programmers than those who did save *smaller revisions*.
- *Programming styles* entered into regression analysis.
- Seeing computer science as a male domain improved performance in programming

## 5. Discussion

In this section we summarize our main findings.

1- Influence of gender on *programming styles*: Our study confirms that males and females differ in *bug-removal-avoidance* style. It might be that, since females are more humble or less confident, they acknowledge their bugs more frequently. However, we could not find any relations between gender and *bottom-up* style as claimed by Cox and Fisher (2006). It might be that, since there are few females (7 females vs. 58 males) in our sample, we could not study gender appropriately.

2- Influence of experience on *programming styles*: Our study confirms the findings of the previous literature in the sense that our styles change with experience (Mayrhauser 1997). In particular, we found that months/years of programming experience affect the *bottom-up* style positively and years at university affect it negatively. Mayrhauser (1997) explains that knowledge of programming languages encourages *bottom-up*. We mainly measured programming experience in coding; therefore, it seems logical that more years of programming experience encourage programmers' use of the *bottom-up* style. Higher years at university means that they have done advanced courses and have more experience with models. Therefore, it seems reasonable that higher-year students do not only rely on the *bottom-up* style. However, we could not find any relations of the *top-down* style. It might be that those higher-year programmers use *top-down* where appropriate, and since our indicators are independent of problem and situation, we could not measure it.

Although the existing research (Vessey 1985, Ko and Uttl 2003) indicates that experience affects *breadth-first* and *depth-first* styles, more experienced programmers use the *breadth-first* style and less experienced ones use the *depth-first* style. We did not find any relations between experience and *breadth-first* or *depth-first* styles. It might be that this relation appears in very different experiences, for example, experts and novices (Vessey 1985) and computer-science and other majors (Ko and Uttl 2003). Another possible explanation is that task difficulty mediates theses relations, because they have been found in the previous studies (Vessey 1985, Ko and Uttl 2003) on difficult tasks.

3-Influence of attitude on *programming styles*: We found that attitude towards programming has a positive effect on *depth-first*, *breadth-first* and *bottom-up* styles. It might be that a positive attitude encourages almost all styles. Another explanation is that individuals with a positive attitude tend to choose positive options in personality items.

4- Influence of personality on *programming styles*: We found that *Openness to Experience* affects *breadth-first* style. It might be that, since high *Openness to Experience* programmers have broader views and see more alternatives, they tend to use *breadth-first* style. Moreover, we found that *Conscientiousness* affects *depth-first* style. It might be that, since high *Conscientiousness* programmers are goal-oriented, they are good at finding fast solutions and, therefore, they tend to use *depth-first* style. As Cox and Fisher (2009) claimed, personality traits might explain the differences in *programming styles*. This finding not only helps programmers to know themselves better but it also helps supervisors to understand the differences between programmers.

5-Influence of human factors on the combination of programming styles: Programmers switch between different styles and one programmer might use different styles. Although we did not focus on the combination of styles, our findings indicate that programmers with a positive attitude towards programming used both *breadth-first* and *depth-first* style. It might be that their positive attitude helps them to switch to the better style according to situation. Moreover, it makes sense that experienced programmers switch among styles when needed. However, we could not find this effect in the current study. It might be that, since we focused on personality and used indicators in measuring default programming styles, we could not find the influence of experience.

5- Influence of human factors on programming performance. Obviously, prior experience in programming and academic achievements are the most important factors for programming performance (Karimi and Wagner 2014). We also found that older programmers show a better performance. It might be that age indicates general experience and, therefore, older programmers have more experience and do the job better. Additionally we found that programmers who see computer science as a male domain show better performance than programmers who do not share this view. It might be that seeing computer science as a male domain motivates programmers to write better code since they feel that they compete against stronger rivals.

6- Influence of *Programming styles* on programming performance. We found that despite influential human factors like experience, some *programming styles* are preferred. This finding helps to improve programmers and programming. In particular, the findings of this research showed that using the *depth-first* style improves performance. Our findings are in contrast to the previous studies (Vessey 1985 and Ko and Uttl 2003) which found that *breadth-first* style has a positive effect on performance and *depth-first* has a negative effect. It might be that task difficulty mediates relations between programming styles and performance. It means that *breadth-first* is not always a good habit and *depth-first* is not also always a bad habit. *Depth-first* programmers are more productive in iterative programming and, therefore, it might be that they have a better self-perception or are more successful in refining their code.

Additionally, we could not find the influence of *bottom-up* on performance in this study although the previous studies (Fisher et al. 2006) did. It might be that *bottom-up* is not always useful and for example depends on task factors like generating or modifying source code. Another explanation is that our indicators for measuring *bottom-up* are not comprehensive.

Moreover, our findings showed that saving *large revisions* increases programming performance. Larger revisions show perfective actions (Hindle et al. 2008) and it might be that programmers who tended to work with larger revisions have less test-time bugs. Another explanation is that smaller revisions missed co-changes and decreased software quality.

6- Indirect influence of human factors on programming performance via *programming styles*. Human factors affect *programming styles* and *programming styles* affect performance. This indirect effect indicates possible causes for the influence of personality and other human factors on programming performance. Using regression analysis to explain possible casual relationships is out of scope of this study. However, we show the casual model of our findings to illustrate the potential impacts of this study (see figure 6).

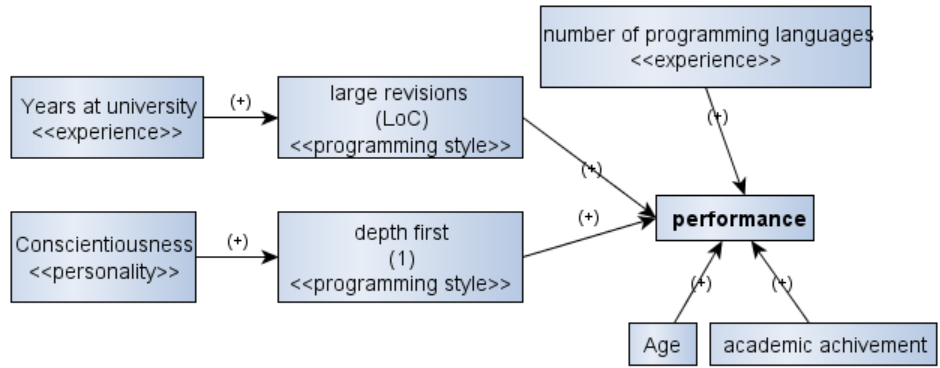

Figure 6: Relationships among programming styles, human factors and programming performance

Figure 6 indicates that the effect of *Conscientiousness* on programming performance is indirect via *depth-first* style. In other words, since *Conscientious* programmers use *depth-first* style, their performance is better. Moreover, figure 6 indicates that the effect of *years at university* on programming performance is indirect via *large-revision*, since programmers with *higher years at university* do not save small revisions have better programming performance.

7-*Programming styles* help to give personal recommendations to programmers. We can use our findings to analyze strengths and weaknesses in programmers and give personal recommendations to them. For example, we recommend:

- To programmers with *low Openness to Experience*: Invest more time in finding alternatives to practice *breadth-first* style.
- To programmers with *low Conscientiousness*: Focus on the goal and find simple solutions to practice *depth-first* style.
- To programmers in low years at university: Be careful about the revisions you save. Avoid small and incomplete revisions.

8- Influence of *programming styles* on team performance. In this study we focused on individual performance in programming. However, the participants worked in teams and there are studies about the influence of individual differences on team success. For example, Goral and Lam (2004) found that teams with *extraverted* programmers show better performance and Acuna et al. (2009) also found that *Extraversion* positively correlated with team success. It is because *extraverted* individuals are more social and enjoy team work. Salleh et al. (2009, 2010a, 2010b, 2011, 2013) conducted a family of experiments and found that paired students at higher levels of *Openness to Experience* got better academic achievements. This is because programmers with higher levels of *Openness to Experience* not only make more suggestions but also build on the ideas of other members which result in higher levels of knowledge sharing and better academic achievement.

We further analyzed the data to investigate which personality factors and *programming styles* affect team success. To do so, we used the average *grade* among team members as team success. We computed min, max and average of all independent variables including personality factors and *programming styles* and used number-of-females to study the effect of gender in teams.

Doing correlations analysis on 14[2] teams we found that *mean-Agreeableness* and *max-Agreeableness* affect team performance positively. *Agreeableness* is important in team work (Howard and Howard 2009) and it seems logical that it increases team performance. We could not find the effect of *Extraversion* in this study, maybe because in this study grade reflects individual performance. We also could not find the effect of *Openness to Experience* in this study. It might be that, since in our study grade does not reflect the learning outcome but the quality of the project, the influence of *Openness to Experience* was not apparent.

Moreover, we found that *min-depth-first.1* and *max-depth-first.1* affect team success. It means using *depth-first* with all team members is useful for team success. This seems logical since *depth-first* programmers do not waste time on reflecting and systematic work and, hence, they share work as soon as possible.

9- Influence of human factors on work schedule. Work schedules depend on task and situational factors like deadlines. However, it was interesting to monitor commit times and study the influence of personality on it. We extracted work percentage at night and weekends and correlated them with human factors.

We found that *Conscientiousness* and *Emotional Stability* positively correlated with *working at weekends* and *age* and *years at university* negatively correlated with working at weekends. Moreover, *Emotional Stability*, *age* and *Conscientiousness* entered into regression analysis. It seems reasonable that *Emotionally Stable* programmers postpone their tasks to weekends. Since younger programmers here took a one-semester basic software project course, it might be that they are under more pressure to meet deadlines and, therefore, work on weekends. Moreover, since Higher *Conscientious* programmers are goal-oriented, they are more motivated to work on weekends.

Additionally, we found that *attitude towards usefulness of programming* affects *working at night* positively. It seems logical that programmers who work at midnight enjoy programming.

## 6. Limitation

First, we used a self-developed questionnaire to operationalize some of the programming styles. There is the threat that questions do not reflect programming styles. We mitigated this threat by relying on the previous literature for formulating the indicators and piloting and revising the questionnaire several times. Moreover, the correlation among indicators shows they are not cohesive. Sometimes the correlations between indicators of different styles are stronger than correlations between indicators of the same style. Therefore, we considered and correlated each indicator separately.

---

We included teams with at least 3 team members in this part of analysis [2]

Second, we used a self-assessed survey to measure some of the programming styles. There is the threat that some respondents provide inaccurate or untrue answers. We mitigated this threat by relying on volunteers who were interested in the research results and in getting personal feedback. Furthermore, the research results were confidential and did not affect their grades. Moreover, we collected several objective programming styles and working schedules from software repositories and the conclusions are therefore quite well supported.

Third, although the personality questionnaire is standard, we found several significant correlations between personality factors. Since significant correlations between the Five Factors have been observed frequently in the Five Factor Manuals (McCrae and Costa 1992 for example), this does not invalidate our findings.

Fourth, more than half of the participants (37 out of 65) are in the second year of their studies. There is the threat that such students are likely to be still at the formative stage of developing their programming style. But generally, the experience of German students in Computer Science is not limited to university and, for example, 22 out of these 37 second-year students in our sample had at least 3 years of programming experience and also 7 students out of 15 less experienced second-year students knew at least 3 programming languages. Therefore, we believe that the programming experience in our sample is not limited. Besides that, we did not ask for the programming styles directly but used basic indicators instead and reviewed and piloted the questions with the help of participants' supervisors who confirmed that the questions made sense for the participants.

Fifth, the participants of this study were student programmers and our results might not be true for the professional population. This result as just conjectures about the styles of professional programmers was interesting for us. Moreover, this result is also interesting for computer science students and their supervisors. We wanted to give them a clue about how their habits are different and how these differences may influence their performance. Therefore, we are able to assist prospective professional programmers by identifying their potential weak points and talents early on.

Sixth, our participants were not in the same year of their education, had different programming experience levels and worked on different projects. Since we have different experience levels and different projects in industry as well, this sample can still represent prospective software engineers.

Seventh, like many studies of this kind, this study suffers from volunteer bias; people who are willing to volunteer their opinion may differ from those who decline to take part.

Finally, we did a correlational study and although personality exists prior to behaviour, the correlation of personality (and other behaviour) still does not reflect cause and effect (Cloninger 2004). Maybe there are some other variables which affect behaviour and correlate with personality (for example motivation).

## 7. Future Work

In this study, we extracted one programming style from software repositories and devised a questionnaire to assess five other programming styles. Neither our mining nor our questionnaire is comprehensive, further research is needed to classify existing programming styles and determine new ones. This can be done by reviewing literature, observing programmers and more mining of software repositories. How to assess programming styles and whether or not assessments are valid are also open challenges.

We did investigate single programming styles. Programmers might switch between styles and each programmer can use multiple styles. Future research is needed to investigate combinations of programming styles and study the influence of human factors like experience on them.

In this study we investigated effects and relations using correlational analysis with Pearson's correlation coefficient and linear regression. Other techniques like association rule mining might discover other interesting relations. Moreover, future studies need to evaluate prediction to help in programmer selection. Moreover, in this study we focused on individual styles and individual performance. Future studies need to investigate different styles in team settings. Additionally, in this study we mainly investigated experience and personality. Future studies should consider other important factors like cognitive styles (Witkin et al. 1967).

There are studies on the implications of programming styles. For example, Wang and Arisholm (2008) found that fault probability in hard-first and easy-first task orders is different and the design approach affects it. It means regarding the design approach, we can prevent a bad programming style. However, the implication of programming styles is not clear yet. Future research needs to find out in which situations and for which tasks a certain programming style is appropriate.

In this study we investigated the programming behaviour of 65 student programmers at the University of Stuttgart in Germany. Future studies should be conducted on larger sample sizes, with professionals and in different locations to strengthen the results.

## 8. Conclusion

The research problem addressed in this paper is to analyze some programming styles, which Fisher and Cox (2009) believe are influenced by personality factors. Looking to validate their claim, we analyzed the relationships between personality factors and programming styles. Moreover, we studied the influence of personality factors and programming styles on programming performance to understand potential benefits of analyzing programming styles.

For that, this paper presents two correlational analyses. In the first study, we analyzed the relations between age, gender, GPA, experience, personality and attitude on one hand and programming styles on the other. In the second analysis, we investigated the relations between all factors, human factors and programming styles, and programming performance. Specifically, we measured programming performance using the quality of code, project and programmer.

We considered all five personality factors, *Openness to Experience*, *Conscientiousness*, *Extraversion*, *Agreeableness*, and *Emotional Stability* and studied the following programming styles: *top-down*, *bottom-up*, *breadth-first*, *depth-first*, *bug-removal-avoidance* and *large-revisions*.

We analyzed the findings and reached the following results. First, we confirm the relationship between personality factors and programming styles. Programmers with *high Conscientiousness* tended to use *depth-first* style. *Conscientiousness* encourages a clear view of the goal and eases opportunistic strategies in programming. Additionally, we can say that programmers with *high Openness to Experience* tended to use a *breadth-first* style. *Openness to Experience* supports seeing alternatives and motivates a systematic style in programming.

Second, we reconfirm that programming styles influence performance in programming. Specifically, we found that *depth-first* and *large-revisions* are good styles. *Depth-first* programmers might be more productive in iterative programming. Additionally, it might be that programmers who tend to do *small-revisions* do not invest enough time in perfective actions and may miss co-changes.

Finally, we conclude that studying programming styles has the following advantages. (1) Programming styles help understand the influence of personality in programming. For example, despite organizational psychology's results on the influence of *Conscientiousness* (Barrick and Mount 1991), we did not see the direct influence of *Conscientiousness* on performance in programming. However, we can say that *Conscientiousness* encourages *depth-first* style which is beneficial in programming. (2) Using programming styles we are able to provide personal recommendations to programmers. For example, based on this research's results, we recommend *low-Conscientious* programmers to examine opportunistic strategies to speed up reaching their goal.

All this knowledge helps us to select better and fitter programmers and train the existing programmers in order to improve the quality of source code, software projects and programmers. We invite future research to consider programming styles in the investigation of personality in computer programming to clarify the influence of personality in programming.

## Acknowledgements

We want to thank all the volunteers who took part in the study, the psychologist advisor of the study: Jan-Paul Leuteritz from Fraunhofer IAO and anonymous reviewers for their effective recommendations. The first author would like to thank the members of the Institute of Software Technology (ISTE), University of Stuttgart, who not only provided an excellent research environment to accomplish the main part of this work there but also helped her in reviewing all the materials and in organizing and conducting data collection. Especially the first author would like to thank Daniel Kulesz for his effective recommendations and all his warm support during data collection and Kornelia Kuhle for the English proofing of all the materials and this paper.

**Appendix A: Questionnaire Items, Answer Alternatives and Descriptive Statistics**

Table A1. General Questions and Descriptive Statistics

| biographical variable | item | answer alternatives and frequencies |
|---|---|---|
| age | 1-What is your age? | **18**,(0) **19**(5), **20**(11), **21**(13), **22**(8), **23**(7), **24**(6), **25 and above**(15) |
| gender | 2-What is your gender? | **Male** (58), **Female**(7) |
| language | 3- Is German your first language, the one in which you do most of your thinking, conversing, creating, writing, and mathematical calculations? | **No** (2), **Yes** (63) |

Table A2. Questions on Programming Experience, academic achievement and Descriptive Statistics

| experience variable | item | answer alternatives and frequencies |
|---|---|---|
| years-of-professional experience | 1-How many months/years of professional programming experience do you have, working full-time excluding school and university work? | **Less than 6 months**(50), **6 months...1 year**(6), **1...3 years**(9) |
| years-of-programming experience | 2-How many years have you been programming including school and university work? | **Less than 1 year**(2),**1 year ... 3 years**(19),**3 years ... 5 years**(22),**More than 5 years**(22) |
| number-of-programming-languages | 3-In which programming languages have you ever contributed code larger than 100 LOC? Pascal, Delphi, Basic, Visual Basic, C, C++, C#, Java, Ada, JavaScript, Python, Others | **1**(7), **2**(8), **3**(16), **4**(15), **5**(10), **6**(6), **7**(1), **8**(2) |
| largest code contribution | 4-How large is the largest code contribution you have ever made individually or as a team member? | **Less than 100 LOC**(3), **100 LOC ... 1000 LOC**(7), **1001 LOC ... 5000 LOC**(21), **5001 LOC ... 10000 LOC**(14), **More than 10001 LOC**(20) |
| academic achievement | 3-How do you roughly estimate your overall (average) grade of all university exams when you are looking at courses you have successfully passed? | **Worse than 3** (weak 5), **2 ... 3**(average 41), **Better than 2**(good 19) |

Table A3. Questions on Programmer ability and Descriptive Statistics

| biographical variable | item | answer alternatives and frequencies |
|---|---|---|
| programmer ability | Which of these options do you usually use to describe yourself in relation to programming? (Please check all applicable answers.) I am an experienced programmer. | **No** (19), **Yes** (66) |
| programmer ability in comparison | Which of these options do you usually use to describe yourself in relation to programming? (Please check all applicable answers.) I am more experienced than my friends in the team. | **No** (48), **Yes** (17) |

Table A4. Questions on Attitude towards Programming and Descriptive Statistics

| attitude scale | items | descriptive statistics |
|---|---|---|

| effective-motivation-in-programming | 1-I enjoy programming.<br>2-I don't understand how some people can spend so much time on writing programs and still seem to enjoy it. | **n** = 31, **min** = 5, **max** = 10, **mean** = 8.94, **Std Deviation** = 1.29 |
|---|---|---|
| confidence-in-learning-programming | 3-I have a lot of self-confidence when it comes to programming.<br>4-For some reason, even though I work hard at it, programming seems unusually difficult to me.<br>5-I am sure I can do advanced work in programming. | **n** = 31, **min** = 4, **max** = 15, **mean** = 11.12, **Std Deviation** = 2.64 |
| computer-science-as-a-male-domain | 6-It makes sense that there are more men than women in computer science.<br>7-Females are as good as males at computer science. | **n** = 31, **min** = 3, **max** = 10, **mean** = 7.51, **Std Deviation** = 3.90 |
| attitude-towards-success-in-programming | 8-Being first in a programming competition would make me proud.<br>9-People would think I was some kind of nerd if I am among the outstanding students in developing software projects. | **n** = 31, **min** = 2, **max** = 10, **mean** = 7.45, **Std Deviation** = 1.82 |
| usefulness-of-programming | 10-Programming is a worthwhile and useful skill to have. | **n** = 31, **min** = 3, **max** = 5, **mean** = 4.64, **Std Deviation** = .55 |

Table A5. Personality factors and Descriptive Statistics on Personality Factors

| personality factor | descriptive statistics |
|---|---|
| Openness to experience | **n** = 65, **min** = 25, **max** = 49, **mean** = 38.41, **Std Deviation** = 5.00 |
| Conscientiousness | **n** = 65, **min** = 25, **max** = 46, **mean** = 36.69, **Std Deviation** = 4.85 |
| Extraversion | **n** = 65, **min** = 16, **max** = 47, **mean** = 35.41, **Std Deviation** = 6.79 |
| Agreeableness | **n** = 65, **min** = 25, **max** = 47, **mean** = 38.2, **Std Deviation** = 4.77 |
| Emotional Stability | **n** = 65, **min** = 21, **max** = 50, **mean** = 34.61, **Std Deviation** = 6.73 |

Table A6. Questions on Programming Styles Indicators and Descriptive Statistics

| programming style | item | frequencies and mean |
|---|---|---|
| bug-removal-avoidance | 3-I usually compile my program only after writing a large amount of code. | **No**(39), **Yes**(26) **mean** = .4 |
| depth-first-1 | 5-Imagine you are fixing a bug in your friend's program. How often do you use this strategy for finding a clue about the actual problem? I take a chance and navigate different parts of code to find a clue. | **No**(36), **Yes**(29) **mean** = .44 |
| depth-first-2 | 6-Imagine you reach the first clue. How often do you use this strategy in relation to the first clues about the actual problem?<br>I evaluate the clue: fix the bug and then test the program. | **No**(19), **Yes**(46) **mean** = .71 |
| breadth-first-1 | 7-Imagine you are fixing a bug in your friend's program. How often do you use this strategy for finding a clue about the actual problem? I read and understand the code to find a clue. | **No**(12), **Yes**(53) **mean** = .82 |
| breadth-first -2 | 8-Imagine you reach the first clue. How often do you use this strategy in relation to the first clues about the actual problem?<br>I do not trust the first clues immediately. I keep reading and understanding the code to make sure. | **No**(44), **Yes**(21) **mean** = .32 |
| bottom-up-1 | 9-Imagine you did some maintenance tasks on your friend's program. What comes to your mind two hours later when you remember the code? I remember the exact name of methods. | **No**(44), **Yes**(21) **mean** = .32 |
| bottom-up-2 | 10-I remember the approximate location of methods. | **No**(14), **Yes**(51) **mean** = .78 |
| top-down-1 | 11-I roughly remember the functionality of methods. | **No**(14), **Yes**(51) **mean** = .78 |
| top-down -2 | 12-I exactly remember how large methods were. | **No**(49), **Yes**(16) **mean** = .25 |